\documentclass{siamltex1213}

\usepackage{graphics}

\usepackage{amsmath} 
\usepackage{amsfonts} 
\usepackage{xspace} 

\newcommand{\whp}{\mbox{w.h.p.}\xspace}
\newcommand{\Whp}{\mbox{W.h.p.}\xspace}

\newtheorem{thmObservation}{Observation}[theorem]
\newtheorem{lemmaObservation}{Observation}[theorem]

\newcommand{\eps}{\epsilon}
\newcommand{\tran}{^{\mathrm{\sf \scriptscriptstyle T}}}

\newcommand{\ney}[2]{#1}

\newcommand{\V}{V\!}
\newcommand{\Vmax}[1]{\V_{\max} (#1)}
\newcommand{\Vmin}[1]{\V_{\min} (#1)}
\newcommand{\Smax}[1]{S^*_{\max} (A,#1)}
\newcommand{\Smin}[1]{S^*_{\min} (A,#1)}
\renewcommand{\Smax}[1]{S^*_{\max} (#1)}
\renewcommand{\Smin}[1]{S^*_{\min} (#1)}

\newcommand{\set}[1]{\{#1\}}

\newcommand{\field}[1]{\mathbb{#1}}

\newcommand{\R}{\field{R}}

\newcommand{\Rp}{\R_{+}}

\newcommand{\ones}{{\mathbf 1}}

\newcommand{\LINEL}[4]
{\label{#1}
  \lefteqn{{\parbox{2.4in}{\hfill$#2 #3$}}}
  \text{{\mbox{\hspace*{.7in}}}}
  &\text{{\mbox{\hspace*{1.7in}}}}
  &\text{\parbox[t]{1.76in}{\raggedright\footnotesize #4}}
}
\newcommand{\LINE}[4]
{\label{#1}
  #2 & #3 & \text{\parbox[t]{1.76in}{\raggedright\footnotesize #4}}
}

\title{On the Number of Iterations for {D}antzig-{W}olfe Optimization 
and Packing-Covering Approximation Algorithms\thanks
{A conference version of this paper appeared in the proceedings of IPCO, 1999~\cite{KleinY99}.}}
\author{
  Philip Klein\thanks
{Brown University, Providence, RI{}; 
research partially supported by NSF Grants CCR-9700146
and CCF-0635089.}
  \and
  Neal E. Young\thanks
{University of California, Riverside, CA{};
research partially supported by NSF grants CCR-9720664, CNS-0626912, CCF-0729071, CCF-1117954.}
  }

\begin{document}
\maketitle
\slugger{sicomp}{xxxx}{xx}{x}{x--x}

\begin{abstract}
  We give a lower bound on the iteration complexity 
  of a natural class of Lagrangean-relaxation algorithms
  for approximately solving packing/covering linear programs.
  We show that, 
  given an input with $m$ random 0/1-constraints on $n$ variables,
  with high probability,
  \emph{any} such algorithm requires 
  \(\Omega(\rho \log(m)/\eps^2)\)
  iterations to compute a $(1+\eps)$-approximate solution,
  where $\rho$ is the \emph{width} of the input. 
  The bound is tight for a range of the parameters $(m,n,\rho,\eps)$.

  The algorithms in the class include 
  Dantzig-Wolfe decomposition, 
  Benders' decomposition, 
  Lagrangean relaxation as developed by Held and Karp [1971] for lower-bounding TSP, 
  and many others
  (e.g.~by Plotkin, Shmoys, and Tardos [1988]
  and Grigoriadis and Khachiyan [1996]).
  To prove the bound, we use a discrepancy argument 
  to show an analogous lower bound 
  on the support size of $(1+\epsilon)$-approximate mixed strategies
  for random two-player zero-sum 0/1-matrix games.
\end{abstract}



\section{Background}
We consider a class of algorithms that we call \emph{Dantzig-Wolfe-type} algorithms.  
The class encompasses algorithms from three lines of research.
One line began in 1958 with a method proposed 
by Ford and Fulkerson~\cite{FordF}
for Multicommodity Flow.
Dantzig and Wolfe~\cite{DantzigW}  generalized it as follows.
They suggested decomposing an arbitrary linear program into two sets of constraints, 
as
\[ \min\set{c\tran x :  Ax \geq b,\, x \in P}
\]
where $P$ is a polyhedron, and using an algorithm that solves the
program iteratively.  In each iteration, the algorithm 
performs a single linear optimization over the polyhedron $P$
--- that is, in each iteration, the algorithm chooses a cost vector
$q$, and computes
\[\arg\min\{q\tran x:x\in P\}.\]

This approach, now called \emph{Dantzig-Wolfe decomposition}, 
is especially useful when $P$ is a Cartesian product $P_1 \times \cdots \times P_K$
and linear optimization over $P$ decomposes into independent optimizations over each $P_i$.  

\paragraph{Lagrangean relaxation}  
In 1970, Held and Karp~\cite{HeldK1,HeldK2} proposed a now well-known 
lower bound for Traveling Salesman Tour,
which they formulated (for some $(A,b,c)$) as the mathematical program
\[  \max_u \Big[ u\tran b + \min_{x \in P} (c-u \tran A)x \Big].
\]
Here $P$ is the polyhedron whose vertices are 1-trees (spanning trees plus one edge;
a relaxation of traveling salesman tours).
To compute an approximate solution, they suggested
starting with an arbitrary assignment to $u$, then iterating as follows:
find a minimum-cost 1-tree $T\in P$ with respect to the edge costs $q = c-u\tran A$;
increase $u_v$ for each node $v$ of degree 3 or more in $T$, then repeat.

As in Dantzig-Wolfe decomposition, their algorithm interacts 
with the polyhedron $P$ only by repeatedly
choosing a cost vector $q$ and solving for $T=\arg\min\{q\tran x:x\in P\}$.
The method has been applied to a variety of other problems,
and has come to be known as Lagrangean relaxation.
It turns out to be the \emph{subgradient method}, which dates back to the early sixties.  

\paragraph{Fractional packing and covering}
In 1979, Shapiro~\cite{Shapiro} referred to the ``the correct combination of artistic expertise and luck'' 
needed to make progress in subgradient optimization ---
although Dantzig-Wolfe decomposition and Lagrangean relaxation
could sometimes be proved to converge in the limit, 
in practice, finding a way to compute and use queries 
that gave a reasonable convergence rate was an art.

In contrast, the third line of research provided guaranteed convergence rates.  
In 1990, Shahrokhi and Matula~\cite{ShahrokhiM} gave an approximation algorithm 
for a special case of Multicommodity Flow,
which was improved by Klein, Plotkin, Stein, and Tardos~\cite{KleinPST94}, 
by Leighton et al.~\cite{LeightonMPSTT95}, and others.  
Plotkin, Shmoys, and Tardos~\cite{PlotkinST94} 
generalized it to approximate fractional packing (defined below);
Grigoriadis and Khachiyan obtained similar results independently~\cite{grigoriadis1996coordination}.
Many subsequent algorithms (too many to list here) build on these results,
extending them to fractional covering and to mixed packing/covering,
and improving the convergence bounds in various ways.
Generally, these algorithms are also of the Dantzig-Wolfe type:
in each iteration, they do a single linear optimization over the polyhedron $P$.

This research direction is still active.
Bienstock gives an implementation-oriented, operations-research perspective~\cite{bienstock2002pfm}.  
Arora et al.~give a computer-science perspective, highlighting connections to other fields such as learning theory~\cite{arora2012multiplicative}.
An overview by Todd places them in the context of general linear programming~\cite{todd2002mfl}.

In many applications,
the total time for the algorithm is the number of iterations times the time per iteration.
In most applications, the time per iteration (to solve the subproblem) is large (e.g.~linear or more).
Hence, a main research goal is to find algorithms that take as few iterations as possible.
This paper concerns the following question:
\emph{How many iterations (i.e., linear optimizations over the underlying polyhedron $P$)
  do Dantzig-Wolfe-type algorithms require
  in order to compute approximate solutions to packing and covering problems?}
We give lower bounds
(worst-case and average-case) that match
known worst-case upper bounds for a range of the relevant parameters.

\paragraph{Definition of Dantzig-Wolfe-type algorithms for packing/covering}
We start with a formal definition of packing and covering.

\begin{definition}[fractional packing and covering~\cite{PlotkinST94}]
  An instance of \emph{fractional packing (or fractional covering)}  is a triple $(A, b, P)$, 
  where $A$ is in $\R^{m\times n}$, $b$ is in $\Rp^m$ 
  and $P$ is a polyhedron
  in $\R^n$ such that $Ax\ge 0$ for all $x\in P$.
  A \emph{feasible solution} is any member of the set
  $\{ x \in \R^n : A x \leq b,\, x \in P\}$.
  (For covering, the constraint $Ax\leq b$ is replaced by $Ax\ge b$.)

  If such an $x$ exists, the instance $(A,b,P)$ is called \emph{feasible}.
  A \emph{$(1+\epsilon)$-approximate solution} is an $x\in P$
  such that $Ax \leq (1+\epsilon) b$
  (for covering, such that $Ax \geq b/(1+\eps)$).  
\end{definition}

Informally, a Dantzig-Wolfe-type algorithm, given a packing instance $(A,b,P)$,
computes a $(1+\eps)$-approximate solution,
interacting with $P$ only via linear optimizations of the following form:
\begin{equation}\label{linopt}
  \mbox{\emph{Given some $q\in \Rp^n$, find an $x\in P$ minimizing $q\tran x$}}.
\end{equation}
In our formal model, instead of $P$, the algorithm is given an \emph{optimization oracle} for $P$,
defined as follows.
\begin{definition}[Dantzig-Wolfe-type algorithm for packing]\label{def:dw}
  For any polyhedron $P\subseteq\Rp^n$,
  an \emph{optimization oracle $X_P$ for $P$} 
  is a function $X_P:\Rp^n\rightarrow P$
  such that, for every input $q\in \Rp^n$, the output $x^*=X_P(q)$ satisfies
  $x^*\in P$ and $q\tran x^* = \min\{ q\tran x : x\in P\}$.
  
  An algorithm is of \emph{Dantzig-Wolfe type}
  if, for each triple $(A,b,X_P)$ 
  where $(A,b,P)$ is a packing instance and $X_P$ is an optimization oracle for $P$,
  the algorithm (given input $(A,b,X_P)$)
  either decides correctly that the input $(A,b,P)$ is infeasible,
  or outputs a $(1+\epsilon)$-approximate solution.
  The algorithm accesses $P$ only by linear optimization via $X_P$:
  in each iteration, the algorithm computes one \emph{oracle input} $q\in\Rp^n$,
  then receives the \emph{oracle output} $X_P(q)$\,
\end{definition}
For covering, the definition is the same, with ``$\max$'' replacing ``$\min$''. 

The oracle $X_P$ above models how most Dantzig-Wolfe-type algorithms
in the literature work, and how they are analyzed: 
their analyses show that they finish within the desired time bound
given any optimization oracle $X_P$ for the polyhedron $P$.
This paper studies the limits of such algorithms, or, more precisely, such analyses.
For our lower bounds, all parts of the input $(A,b,X_P)$, including $X_P$,
are chosen by an adversary to the algorithm.
Although the oracle $X_P$ is not completely determined by the polyhedron $P$,
the distinction between $X_P$ and $P$ is a minor technical issue.\footnote
{The value of $X_P(q)$ is determined by the polyhedron $P$ 
  for all oracle inputs $q\in\Rp^n$
  except those that happen to be orthogonal to an edge of $P$,
  for which $\min\{ q\tran x : x \in P\}$ has multiple minima,
  where $X_P(q)$ can break the tie arbitrarily.}

In the Held-Karp computation (for bounding the optimal traveling-salesman tour)
each oracle call $X_P(q)$
reduces to a minimum-spanning-tree computation with edge-weights given by $q$.
For multicommodity-flow problems,
each oracle call typically reduces (depending on the underlying polyhedron) 
to either a shortest-path computation with edge weights given by $q$,
a minimum-cost single-commodity-flow computation with edge costs given by $q$,
or several such computations (one per commodity).

\section{Main result: lower bound on iteration complexity}
Recall our main question:
\emph{how many iterations (i.e., oracle calls) does a Dantzig-Wolfe-type algorithm 
  require in order to compute $(1+\eps)$-approximate solution to a packing and covering problem}?
Each call reveals some information about $P$.
The algorithm must force the oracle to eventually reveal enough information
to determine an $x\in P$ such that $Ax\le (1+\eps)b$.
In the worst case (for an adversarial oracle),
how many calls does an optimal algorithm require?
For fractional packing, the algorithm of~\cite{PlotkinST94}
gives an upper bound of 
$$O(\rho \eps^{-2}\log m),$$  
where $\rho$, the \emph{width} of the input
is $\rho(A,b,P)=\max_{x \in P} \max_i A_i x/b_i$
(where $A_i$ denotes the $i^{th}$ row of $A$).
%
%
%
%
Our main result 
(Theorem~\ref{thm:main})
is a lower bound that matches this upper bound for a range of parameters.
Here is a simplified form of that lower bound:
\begin{corollary}[iteration bound, simple form]\label{cor:main}
  For every $\delta \in (0,1/2)$, 
  there exist positive $k_\delta, c_\delta>0$ such that the following holds.
  For every two integers $m, n \ge k_\delta$
  and every $\rho\ge 2$,
  there exists an input $(A,b,X_P)$  (packing or covering, as desired)
  having
  $m$ constraints, $n$ variables, 
  and width $O(\rho)$,
  with the following property:

  For every $\eps\in(0,1/10)$,
  every deterministic Dantzig-Wolfe-type algorithm,
  and every Las-Vegas-style\footnote
  {An algorithm having zero probability of error.}
  randomized Dantzig-Wolfe-type algorithm,
  requires at least
  \[c_\delta \cdot \min(\rho\, \eps^{-2}\log m,\, m^{1/2-\delta},\, n)\]
  iterations
  to compute a $(1+\eps)$-approximate solution, given input $(A,b,X_P)$.
\end{corollary}
That is, for every $\delta\in(0,1/2)$,
the worst-case iteration complexity of every Dantzig-Wolfe-type algorithm
is at least \(\Omega_\delta\big(\min(\rho\, \eps^{-2}\log m,\, m^{1/2-\delta},\, n)\big)\).
Here we use the notation $\Omega_\delta$ to signify that the constant
factor hidden by the $\Omega$ notation is allowed to depend on
$\delta$ (but no other parameters).

Section~\ref{sec:sketch} sketches the proof idea.
Section~\ref{sec:main} gives a more detailed version (Theorem~\ref{thm:main}) with full proof.
Theorem~\ref{thm:main} shows that in fact
the bound holds with probability $1-O(1/m^2)$ for random inputs drawn from a natural class:
the polyhedron $P$ is the regular $n$-simplex, $P=\{x\in\Rp^n : \sum_i x_i = 1\}$, and the constraint matrix $A$ is a random 0/1 matrix with i.i.d.~entries.  
The resulting problem instance $(A,b,P)$ 
is equivalent to finding an optimal mixed strategy for the column player
of the two-player zero-sum game with payoff matrix $A$.   
(As a packing problem, the instance models the column player being the min player;
as a covering problem, it models the column player being the max player.)
The basic idea of the proof is to prove a corresponding lower bound
on the minimum support size of any $(1+\eps)$-approximate solution $\hat x$,
and then to argue that (for the inputs in question) each iteration 
increases the support size of $\hat x$ by at most 1.

\paragraph{Extending to products of polyhedra}
Following one of the original models for Dantzig-Wolfe decomposition,
many algorithms in the literature specialize 
when the polyhedron $P$ 
is a Cartesian product $P=P_1 \times \cdots \times P_K$ of $K$ polyhedra
and optimization over $P$ decomposes into independent optimizations 
over the individual polyhedra $P_i$.
It is straightforward to extend our lower bound to this model
by making $A$ block-diagonal, thus forcing each subproblem to be solved independently.
Extended in this way, the lower bound shows
that the number of iterations (each optimizing over some individual polyhedron $P_i$)
must be $\Omega(\sum_i \min(\epsilon^{-2} \rho_i \log m_i, m_i^{1/2-\delta}, n_i))$,  
where polyhedron $P_i$ has $n_i$ variables and width $\rho_i$, 
and $A$ has $m_i$ constraints on $P_i$'s variables.
This lower bound matches known upper bounds 
(e.g.~$O(\sum_i \epsilon^{-2} \rho_i \log m_i)$)
for a range of the parameters.

\subsection{Comparison with previous and related works}
Recall the known upper bound of $O(\rho\,\eps^{-2}\log m)$ iterations
in the worst case~(e.g.~\cite{PlotkinST94}).
It follows that the lower bound here is tight for a certain range of the parameters:
roughly, in the regime \(\rho\eps^{-2} \ll \min(\sqrt m, n)\).
This suggests two directions for proving stronger upper bounds.
The first direction
is to look for better upper bounds outside of the regime \(\rho\eps^{-2} \ll \min(\sqrt m, n)\).
A few such bounds are known
(e.g.~$O(\min(\rho, m)\, \eps^{-2} \log m)$ iterations
\cite{garg2007faster,young2001sequential}
and $O(m (\eps^{-2} + \log m))$ iterations
\cite{grigoriadis2001approximate})
but these leave a large gap w.r.t.~any known lower bound.
The second direction is to consider non-Dantzig-Wolfe-type algorithms,
as discussed later.

\paragraph{Dantzig-Wolfe-type algorithms that allow \emph{approximate} oracles}
Many Dantzig-Wolfe-type algorithms in the literature are known to work
even if run with an \emph{approximate} optimization oracle.
Define a  $(1+\epsilon)$-\emph{approximate oracle} to be a function $X'_P:\Rp^n\rightarrow P$
such that, for all $q\in \R^n$, 
\medskip 

\noindent
\centerline{\em the output $x'=X'_P(q)$
  satisfies $x'\in P$  and $q\tran x' \le (1+\eps)\min\{q\tran x : x\in P\}$,}

\medskip 
\noindent
A typical analysis proves a worst-case performance guarantee such as the following:
\emph{for every input $(A,b,X'_P)$ such that $X'_P$ is a
  $(1+\epsilon/10)$-approximate oracle,
  the algorithm computes a correct output using \(O(\rho \log(m)/\eps^2)\)
  oracle calls.}
A common motivation is that approximate oracles can require less time per iteration,
leading to faster total run times. 

Such an algorithm is, formally, of Dantzig-Wolfe-type per Definition~\ref{def:dw}.
(The reason is trivial: 
every exact optimization oracle $X_P$ per Definition~\ref{def:dw}
is also a valid approximate oracle as defined above,
so such an algorithm necessarily works with every exact oracle as well.)
Hence, the lower bounds in Corollary~\ref{cor:main} and Theorem~\ref{thm:main}
apply to every such algorithm.

As we discuss next, our lower bounds imply that to obtain a better upper bound
requires not only (i) an algorithm that uses an optimization oracle
that does something other than pure linear optimization over $P$,
but also (ii) an analysis that makes use of that additional requirement.

\paragraph{Non-Dantzig-Wolfe-type algorithms} 
To obtain better general upper bounds 
for the parameter regime where the lower bound is tight,
one has to consider non-Dantzig-Wolfe-type algorithms.
Indeed, since the appearance of the conference version of this paper~\cite{KleinY99},
researchers~\cite{chudak2005improved,bienstock2006approximating,iyengar2011approximating,nesterov2009fast} have built on the methods of Nesterov~\cite{nesterov2005smooth} (see
also Nemirovsky~\cite{nemirovski2004prox}) to obtain polynomial-time approximation schemes whose running times have
better dependence on $\epsilon$. These algorithms bypass the lower
bound by optimizing nonlinear convex functions instead of linear functions
(or by linear optimization over $P$ but with side constraints).

Bienstock and
Iyengar~\cite{bienstock2006approximating} give
an algorithm that, for a given $\epsilon>0$ and packing input
  $$\{ x \in \R^n : A x \leq b,\, x \in P\}$$
finds a $(1+\epsilon)$-approximate solution by using calls to a convex
quadratic program over a set of the form
$$\set{x\in P: \forall j.~ 0 \leq x_j \leq \lambda}$$
where the value of $\lambda$ can be adjusted by the algorithm in each
iteration.  
Such an algorithm violates the assumption of our lower
bound in two ways: the objective function is nonlinear, and the
optimization takes place not over $P$ but over the intersection of $P$
with a hypercube of specified side-lengths.  Bienstock and Iyengar also give an
algorithm for covering; it similarly violates the assumptions of our lower bound.  
For their algorithms, the number of
iterations is bounded by
$O(\epsilon^{-1} \sqrt{Kn\log m})$, where $K$ is the maximum number of
nonzero elements in any row of $A$.
Each iteration calls the quadratic-programming oracle.

How difficult is convex quadratic programming?  Using the ellipsoid
algorithm (see~\cite {lovasz1987algorithmic,grötschel2011geometric}),
quadratic programming over an $n$-dimensional convex set can be
reduced to a polynomial number of calls to a linear-optimization
oracle for that set.  However, the polynomial is quite large.
Bienstock and Iyengar also show that it suffices to approximate the
convex quadratic objective function by a piecewise linear objective
function.  In either case, the required oracle is generally more expensive
computationally than linear optimization over the original convex set.

Bienstock and Iyengar illustrate their method with an application to
variants of Multicommodity Flow.  Nesterov~\cite{nesterov2009fast}
also gives an approximation algorithm for a variant of Multicommodity
Flow.  In both cases, the number of iterations is proportional to
$\epsilon^{-1}$ instead of $\epsilon^{-2}$.  However, the dependence
of the overall running time on the size of the problem is 
worse, by a factor of at least the number of commodities.  

Chudak and Eleut{\'e}rio build on the techniques of Nesterov to give
an approximation scheme for a linear-programming relaxation of 
Facility Location~\cite{chudak2005improved}.  The running time of
their algorithm is $\tilde O({(nm)}^{3/2}/\eps)$,
where $nm$ is the number of facilities times the number of clients.
In contrast, a Dantzig-Wolfe-type algorithm can be implemented to run
in time $\tilde O(N/\eps^2)$, where $N \le nm$ 
is the input size --- the number of (facility, client) pairs with finite distance
\cite{young2014nearly}.

Iyengar, Phillips, and Stein~\cite{iyengar2011approximating} use the
method of Nesterov to obtain
approximation schemes for certain semidefinite programs.  For problems
previously addressed using the method of Plotkin, Shmoys, and
Tardos~\cite{PlotkinST94}, their running times, while proportional to
$\epsilon^{-1}$, have worse dependence on problem size.

For the important special case when the polyhedron $P$ is the positive orthant
(e.g., problems of the form $\max\{c\tran x : A x \le b, x\ge 0\}$),
a recent breakthrough by Allen-Zhu and Orecchia
runs in $\tilde O(N/\eps)$ time for packing,
or $\tilde O(N/\eps^{1.5})$ time for covering,
where $N$ is the number of non-zeros in the constraint matrix~\cite{allen2014nearly}.
The algorithms are not Dantzig-Wolfe-type algorithms.

\paragraph{Does the regime in which the bound is tight contain interesting problems?}
Recall that the bound is tight in (roughly) the regime \(\rho\eps^{-2} \ll \min(\sqrt m, n)\).
For some interesting classes of problems,
the width $\rho$ is either constant
(for example, zero-sum games with payoffs in $[0,1]$ and value bounded away from 0 and 1)
or a function of $m$ and/or $n$ that grows slowly
(a celebrated recent example is for Maximum Flow in undirected graphs~\cite{ChristianoKMST11},
in which, for $n$-node graphs, the width is $\widetilde O(n^{1/3})$).
``Small width'' problems such as these (with, say, constant $\eps$)
lie in the regime.

\paragraph{Related lower bounds}
\ney{
  Khachiyan~\cite{Khachiyan}
  proves an $\Omega(\epsilon^{-1})$ lower bound
  on the number of iterations to achieve an error of $\epsilon$.
}{
  In 1977, Khachiyan~\cite{Khachiyan} proved a lower bound for this
  model; he showed that for a 
  certain class of inputs, the error decreases at best
  linearly with the number of iterations. (The constant of
  proportionality in his result depended on the matrix $A$ and on the
  initial error.)  This result implies an $\Omega(\epsilon^{-1})$ bound
  on the number of iterations to achieve an error of $\epsilon$.
}
Grigoriadis and Khachiyan~\cite[\S 2.8]{grigoriadis1996coordination} observe that 
for the packing problem ``\emph{find $x\in\Delta^m$ such that $I x\le \ones/m$}''
(where $\Delta^m$ is the $m$-simplex,
$I$ is the identity matrix,
and $\ones$ is the all-ones vector in $\R^m$)
any 0.5-approximate solution $x$ has to have support of size at least $m/2$,
and that this gives an $m/2$ lower bound on the number of oracle calls
for any Dantzig-Wolfe-type algorithm to return a 0.5-approximate solution.
(Consider also that the \emph{covering} problem
 ``\emph{find $x\in\Delta^m$ such that $I x\ge \ones/m$}''
requires at least $m$ iterations to return \emph{any} approximate solution.)
These inputs have large width, $\Theta(m)$, 
complementing our lower bound.

Grigoriadis and Khachiyan~\cite[\S 3.3]{grigoriadis1996coordination}
generalize their observation above to give a lower bound
on the number of calls required by any algorithm in a class they 
call \emph{restricted price-directed-decomposition} (PDD).
Their model, different from the one studied here,
focuses on product-of-polyhedra packing inputs,
of the form $x\in P=P_1\times P_2\times \cdots \times P_K$ and $Ax\le b$.
In each iteration, the algorithm computes a single vector $y$
and the oracle returns an $x \in \widehat P$ minimizing $(y\tran A) x$,
where $\widehat P = \{x \in P : \forall j.~ x_j \le \mu_j \}$, 
for some vector $\mu$ (subject, crucially, to restrictions on $\mu$).
They show that any such algorithm must use at least $\min(m,k)/\mathrm{polylog}\, m$
iterations to compute a $0.5$-approximate solution.

Freund and Schapire~\cite{FreundS97},
in independent work in the context of learning theory,
prove a lower bound on the net ``regret'' of any adaptive strategy
that plays repeated zero-sum games against an adversary.
Their proof is based on repeated random games.
They study a wider class of problems (giving the adversary more power),
so their lower bound does not apply to Dantzig-Wolfe-type algorithms
as defined here.

\paragraph{Sublinear-time randomized algorithms for explicit packing and covering}
In the special case of two-player zero-sum games with payoff matrix $A$
where each payoff $A_{ij}$ is in $[0,1]$ 
randomized algorithms can compute solutions with additive error $\eps$
in sublinear time
\cite{grigoriadis1995sublinear}
(see also~\cite{Koufogiannakis13Nearly}).
Deterministic algorithms cannot~\cite{grigoriadis1995sublinear}.

\section{Small-support mixed strategies for zero-sum games} 
\label{sec:game_intro}

To prove the lower bound on iteration complexity, we prove an analogous lower bound 
(Theorem~\ref{thm:game})
on the minimum support size\footnote{The support of $x$ is the set $\{j : x_j \ne 0\}$.}
of any $(1+\epsilon)$-approximate mixed strategy $x$
for two-player zero-sum games.\footnote
{A \emph{mixed strategy} for the column player of $A$ is an $x\in\Delta^n$,
where $\Delta^n = \{x \in \Rp^n : \sum_j x_j = 1\}$ is the regular $n$-simplex.
The \emph{expected payoff (or value)} of $x$ (for {\sc max} as the column player) is $\min_i A_i x$.
The value of the \emph{game} $A$ (with {\sc max} as the column player),
is $\max_{x\in\Delta^n} \min_i A_i x$, i.e.,
the maximum expected payoff of any mixed strategy.
With {\sc min} as the column player, the value of the game is $\min_{x\in\Delta^n} \max_i A_i x$.
A \emph{$(1+\epsilon)$-approximate} mixed strategy $x$ 
is one whose expected payoff is within a factor of $1+\epsilon$ of the value of the game.}
Here is a simplified form of the support-size lower bound:
\begin{corollary}[support bound, simple form]\label{cor:game}
  For every $\delta\in(0,1/2)$, 
  there exist  $k_\delta>0, c_\delta>0$ such that,
  for every two integers $m, n \ge k_\delta$ and every $p\in(0,1/2)$,
  there exists a two-player zero-sum matrix game $A$ with $m$ rows, $n$ columns, 
  and value $\Omega(p)$, having the following property:

  For every $\eps\in(0,1/10)$,
  every $(1+\eps)$-approximate mixed strategy for the column player of $A$
  (as either the max player or the min player)
  has support size at least
  \[c_\delta\cdot \min(p^{-1}\, \eps^{-2}\log m,\, m^{1/2-\delta},\, n).\]
\end{corollary}
 %
Section~\ref{sec:sketch} sketches the proof idea.
Section~\ref{sec:game} fully proves a more detailed version (Theorem~\ref{thm:game}),
showing that in fact the bound holds with probability $1-O(1/m^2)$ 
when the payoff matrix $A$ is a random 0/1 matrix with i.i.d.~entries.  

\paragraph{Matching upper bound}
The lower bound in Theorem~\ref{thm:game}
matches (up to constant factors) a previous small-support upper bound by Lipton and Young
\cite{lipton1994simple}:
\emph{For every two-player zero-sum game with payoffs in $[0,1]$ and value $p$,
  each player has a $(1+\eps)$-approximate mixed strategy 
  with support of size \emph{at most} $O(p^{-1}\eps^{-2} \log m)$,
  where $m$ is the number of pure strategies available to the opponent}
The proof is simple.\footnote
{Consider a mixed strategy that plays
  a pure strategy chosen uniformly from a multiset $S$ of $s$ pure strategies,
  where $S$ is formed by sampling $s$ times i.i.d.~from the optimal mixed strategy.
  Use a standard Chernoff bound and the union bound
  to show that this mixed strategy has the desired properties with positive probability.}
Derandomizing the proof via the method of conditional probabilities
gives a Dantzig-Wolfe-type algorithm to compute the $(1+\eps)$-approximate strategy
using $O(p^{-1}\, \eps^{-2}\log m)$ oracle calls~\cite{Young95}.

In the context of Nash equilibria,
similar small-support upper bounds have subsequently been shown
and used for algorithmic upper bounds
(e.g.~\cite{lipton2003playing,feder2007approximating,hazan2011hard}).

\section{Proof ideas}\label{sec:sketch}

This section sketches how a support-size bound (Corollary~\ref{cor:game}) 
implies an iteration-complexity bound (Corollary~\ref{cor:main}),
and how we prove a support-size bound such as Corollary~\ref{cor:game}.
See \S~\ref{sec:game} and \S~\ref{sec:main}
for the more detailed theorems that imply these corollaries, 
with detailed proofs based on the ideas sketched here.

\paragraph{How a support-size bound implies an iteration bound}
We sketch the idea for packing.  The idea also works for covering.
Fix the parameters $m$, $n$, $\rho$ as in Corollary~\ref{cor:main}.
Let probability $p=1/\rho$.
Let $A$ be the $m\times n$ payoff matrix for any zero-sum game 
with the properties described in Corollary~\ref{cor:game}.

Let $\Vmin{A}$ denote the value of the game with {\sc min} (the min player) as the column player.
Let {\sf packing}$(A)$ 
denote the packing problem $(A, b, \Delta^n)$, where each $b_i = \Vmin A$
and $\Delta^n = \{x \in \Rp^n : \sum_j x_j = 1\}$ is the simplex.
This is equivalent to the zero-sum game with payoff matrix $A$ and {\sc min} as the column player.
Via this equivalence, any $(1+\eps)$-approximate solution $\hat x$ for {\sf packing}$(A)$
is also a $(1+\eps)$-approximate mixed strategy for {\sc min} as the column player of the game.
Assuming Corollary~\ref{cor:game}, any such solution $\hat x$
must have support of size $\Omega_\delta(\min(\rho\,\eps^{-2}\log m, \, m^{1/2-\delta},\, n))$,
where $\rho=1/p$.

Whenever the Dantzig-Wolfe-type algorithm queries the oracle for $\Delta^n$,
the oracle can respond to the query $q$ with a vertex of $\Delta^n$.
Each such vertex has just one non-zero coordinate.
For the algorithm to be correct, the final solution $\hat x$ 
must be a convex combination of these vertices, 
so the number of queries must be at least the size of the support of $\hat x$.
To finish, note that  the width of {\sf packing}$(A)$ is $O(\rho)$ 
because the width is $1/V_{\min}(A) = 1/\Omega(p)$.

\paragraph{Proving the support-size bounds (e.g.~Corollary~\ref{cor:game})} 
We sketch a proof of Corollary~\ref{cor:game}
when the column player is {\sc min}.  (The other case is similar.)
Fix the parameters $m$, $n$, $p$ as in Corollary~\ref{cor:game}.
Let $\ell = p^{-1} \eps^{-2} \log m$ be the desired lower bound.

Take $A$ to be a random 0/1 matrix with i.i.d.~entries, 
where each entry $A_{ij}$ is 1 with probability $p$.
\Whp, the value of $A$ is at least $(1-\eps)p$.
(This is easily proven by considering {\sc max}'s uniform mixed strategy.)
Now consider any subgame $B$ of $A$ induced by just $\ell$ columns.
The subgame $B$ is highly skewed 
--- there are many more rows for {\sc max} than columns for {\sc min} ---
so, by a discrepancy argument, \whp, 
the value of $B$ is high: \emph{at least} $(1+3\eps)p$.
(Here is a sketch of the discrepancy argument. 
$B$ is a random 0/1 matrix where each entry is 1 with probability $p$.
Since the number of rows $m$ is much higher than the number of columns $\ell$,
\whp $B$ has a substantial number of rows that have a relatively large number
--- at least $(1+5\eps)p\ell$ --- of ones,
and, \whp, if {\sc max} (the row player) plays uniformly on just these rows,
{\sc max} guarantees a payoff of at least $(1+3\eps)p$ for the subgame $B$.)

Then subgame $B$ has value at least $(1+3\eps)p$,
while $A$ has value at most $(1+\eps)p$.
Since ${(1+\eps)}^2 p < (1+3\eps) p$,
no $(1+\eps)$-approximate mixed strategy $\hat x$ can be supported by just the columns of $B$.
By a union bound over the $n \choose \ell$ submatrices $B$ with $\ell$ columns,
\whp, there is no such $B$ that can support any $(1+\eps)$-approximate mixed strategy $\hat x$,
in which case there is no $(1+\eps)$-approximate strategy $\hat x$ with support of size $\ell$.


This yields the corollary for any single $\eps \in (0,1)$.
To complete the argument, we extend the bound to all $\eps\in(0,1)$ simultaneously (for the given $A$)
by applying the single-$\eps$ case for $\eps$ in a geometrically increasing sequence
$\{\eps_0, 2\eps_0, 4\eps_2, \ldots, 1/10\}$,
then appealing to monotonicity for the remaining $\eps$.

\section{Theorem~\ref{thm:game} (support bound)}\label{sec:game}







In the rest of this section, we state and prove Theorem~\ref{thm:game}.
Theorem~\ref{thm:game} implies Corollary~\ref{cor:game}.
We first give a few self-contained utility lemmas.
The first is a standard Chernoff bound, which we give without proof.

\begin{lemma}[Chernoff bound]\label{lemma:chernoff}
  Let $X$ be the average of $t$ independent, 0/1 random variables, 
  each with expectation $p\in(0, 1]$.  For every $\eps\in(0, 1]$,

  \smallskip 

  \noindent\emph{~(i)}
    {\(\displaystyle
      \Pr[X\le (1-\eps)p]
      ~\le\,
      \exp\big({-\eps^2 pt/3}),
      \)}

    \smallskip 

    \noindent\emph{(ii)}
    {\(\displaystyle
      \Pr[X\ge (1+\eps)p]
      ~\ge\, 
      \exp\big({-\eps^2 pt/3}).
      \)}
    \smallskip 
\end{lemma}

The next utility lemma states that the Chernoff bound above
is tight up to constant factors in the exponent,
as long as the bound is below $1/e$.
That the Chernoff bound is tight (in most cases) is standard folklore.

\newcommand{\chernoffTight}{
      Let $X$ be the average of $s$ independent, 0/1 random variables (r.v.).
      For every $\eps\in(0, 1/2]$ and $p\in(0, 1/2]$,  if $\eps^2 p s \ge 3$,

      \medskip
      \noindent{\emph{~(i)}}
      If each r.v.\ is 1 with probability at most $p$, then
      {\(\displaystyle
        \Pr[X\le (1-\eps)p]
        ~\ge~ 
        \exp\big({-9\eps^2 p s}).
        \)}

      \medskip
      \noindent{\emph{(ii)}}
      If each r.v.\ is 1 with probability at least $p$, then
      {\(\displaystyle
        \Pr[X\ge (1+\eps)p]
        ~\ge~ 
        \exp\big({-9\eps^2 p s}).
        \)}
  }
\newcounter{tightnesssection}
\setcounter{tightnesssection}{\value{section}}
\newcounter{tightnesscounter}
\setcounter{tightnesscounter}{\value{theorem}}
\begin{lemma}[tightness of Chernoff bound]\label{lemma:tightness}
  \chernoffTight
\end{lemma}

\smallskip

A detailed proof is in the appendix. 

\smallskip

The third utility lemma leverages the Chernoff bound to give straightforward bounds
on the likely value of random matrix games.
Note that independence of the entries of the matrix is assumed only within each individual row.

\begin{lemma}[naive bounds on $\Vmax{}$ and $\Vmin{}$]\label{lemma:naive}
  Let $M$ be a random 0/1 $r\times c$ payoff matrix
  such that within each row of $M$ the entries are independent.
  Let $\eps,p\in(0, 1]$.
  
  \medskip
  \noindent{\emph{~(i)}}
  If each entry of $M$ is 1 with probability at least $p$, then
  \[
  \Pr_M[\Vmin{M\tran} \le (1-\eps)p] ~=~\Pr_M[\Vmax{M} \le (1-\eps)p] ~\le~ r\,\exp({-c\,\eps^2 p/3}).\]

  \medskip
  \noindent{\emph{(ii)}}
  If each entry of $M$ is 1 with probability at most $p$, then
  \[\Pr_M[\Vmax{M\tran} \ge (1+\eps)p] ~=~\Pr_M[\Vmin{M} \ge (1+\eps)p] ~\le~ r\,\exp({-c\,\eps^2 p/3}).\]
 \end{lemma}
 \begin{proof}
   \emph{(i)}
  The equality in (i) holds because, by von Neumann's min-max theorem (strong LP duality),
  $\Vmin{M\tran} = \Vmax M$.
  We prove the inequality.
  {\sc Max} can play a uniform mixed strategy on the $c$ columns.
   By the Chernoff bound, 
   the probability that any given row then gives {\sc min}
   expected payoff less than $(1-\eps)p$ is at most $\exp({-\eps^2 p c/3})$.
   By the union bound, the probability that \emph{any} of the $r$ rows gives {\sc min}
   expected payoff less than $(1-\eps)p$ is at most $r\exp({-\eps^2 p c/3})$.

   \medskip
   \noindent
   \emph{(ii)} Similar ({\sc min} can play a uniform mixed strategy on the $c$ columns).
\end{proof}

The next lemma uses the discrepancy argument outlined in the proof sketch
in Section~\ref{sec:sketch}
to quantify the disadvantage to the column player 
for playing a random game with many fewer columns than rows.
The reader may wish to review
Fig.~\ref{fig:key} for the notation.

\begin{lemma}[skewed game 1]\label{lemma:skewed}
  Let $B$ be a random 0/1 $m\times s$ payoff matrix 
  whose entries are i.i.d., each being 1 with probability $p\in(0, 1/2]$.
  Let $\eps\in(0, 1/10]$.  Assume that $\eps^2 ps \ge 1$.
  Then, for $t = m\,\exp({-250\eps^2ps})$,
  and $\beta=s\exp({-\eps^2 t p/15})$,
  \medskip

  \noindent
  {\emph{~(i)}}
  {\(\displaystyle \Pr_B[\Vmax B \ge (1-3\eps) p] ~\le~2\beta\),}

  \medskip
  \noindent
  {\emph{(ii)}}
  {\(\displaystyle \Pr_B[\Vmin B \le (1+3\eps) p] ~\le~ 2\beta\).}

\end{lemma}
\smallskip

(When we apply the bound, $s$ will be chosen so that $t$ is large and $\beta$ is small.)

\begin{proof}
  \emph{(i)}
  Let $D$ be the submatrix formed by the $\lceil t/2\rceil$ rows of $B$ that have the fewest ones,
  as shown in Fig.~\ref{fig:BD}.
  Say that a row of $B$ is \emph{deviant} if the average of its entries is at most $p'=(1-5\eps)p$.
  We claim that \emph{the probability that $D$ has a non-deviant row is at most $\beta$.}
  \begin{figure}
    \centering
    \includegraphics[height=1.25in]{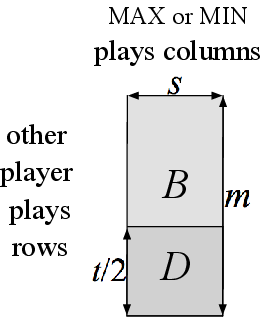}
    \caption{Given skewed matrix $B$, submatrix $D$
      contains the $t/2$ rows with the most (or least) 1's.
      By playing uniformly on the rows of $D$,
      {\sc max} (or {\sc min}) forces value at most $(1-3\eps)p$ 
      (or at least $(1+3\eps)p$) \whp.
    }\label{fig:BD}
  \end{figure}

  To prove the claim, let random variable $d$ be the number of deviant rows in $B$.
  By Lemma~\ref{lemma:tightness} (tightness of the Chernoff bound,
  with $\eps' = 5\eps$,
  using here the assumption $\eps^2ps\ge 1$ and that $\eps\le 1/10$),
  the probability that a given row of $B$ is deviant
  is at least $\exp({-9{(5\eps)}^2ps}) \ge t/m$ (by the choice of $t$).
  Thus, by the choice of $t$, the expected number of deviant rows is at least $t$.
  Since the rows of $B$ are independent,
  by the Chernoff bound (with $\eps=1/2$),
  the probability that $d\le t/2$ 
  is at most $\exp({-{(1/2)}^2 t/3}) = \exp({-t/12})$ which
  (using $\eps\le 1/10$, $p\le 1/2$, and $s\ge 1$) is less than $\beta = s\exp({-\eps^2 t p/15})$.
  This proves the claim, because if $d>t/2$, then all rows in $D$ are deviant.

  Conditioned on all $\lceil t/2 \rceil$ rows in $D$ being deviant,
  within each row of $D$,  by symmetry,
  the probability that any given entry equals 1 is at most $p'=(1-5\eps)p$.
  Also, within any column of $D$ the entries are independent.
  Thus, Lemma~\ref{lemma:naive} part (ii)
  (the naive bounds) applied to the value of the \emph{transpose}, $\Vmin{D\tran}$, implies that
  $\Pr_B[\Vmin{D\tran} \ge (1+\eps)p' \,|\, \mbox{\emph{all rows of $D$ are deviant}}]$
  is at most $s\exp({-(t/2)\eps^2p'/3})$,
  which (using $\eps\le 1/10$ and the choice of $p'$) is at most $\beta=s\exp({-\eps^2 t p/15})$.

  The latter bound and previous claim imply that, unconditionally,
  $\Pr_B[\Vmin{D\tran} \ge (1+\eps)p']$
  is at most $\beta + (1-\beta)\beta$,
  which is less than $2\beta$.
  
  By von Neumann's min-max theorem
  $\Vmax D = \Vmin {D\tran}$.
  Since $D$ consists of a subset of $B$'s rows,
  and {\sc min} is the row player,
  $\Vmax B \le \Vmax D$.
  Transitively,  $\Vmax B \le \Vmin {D\tran}$.
  With the preceding paragraph, this implies that
  $\Pr_B[ \Vmax B \ge (1+\eps)p']$ is at most $2\beta$.
  To finish, note that $(1+\eps)p' = (1+\eps)(1-5\eps)p \le (1-3\eps)p$, as $\eps\le 1/5$.

  \medskip\noindent
  \emph{(Part ii)}
  Say that a row of $B$ is deviant if the average of its entries is at \emph{least} $p'=(1+5\eps)p$.
  Let $D$ be the $\lceil t/2\rceil$ rows of $B$ with the \emph{most} ones.
  Now proceed exactly as in part (i).
  To finish, note that $(1-\eps)p' = (1-\eps)(1+5\eps)p \ge (1+3\eps)p$, as $\eps\le 1/5$.
\end{proof}

\smallskip

We use Lemma~\ref{lemma:skewed} only to prove the next lemma,
which just specializes it to a convenient choice of $s$ (the number of columns).
Namely, we take $s = \lfloor 4\ell \rfloor$, 
where $\ell= \delta\, p^{-1}\eps^{-2}\ln(m)/\,1000$ 
is the lower bound we will seek later.

\begin{lemma}[skewed game 2]\label{lemma:skewed2}
  Let $B$ be a random $m\times s$ 0/1 payoff matrix whose entries are i.i.d.,
  each entry being 1 with the same probability $p\in(0, 1/2]$.
  Let $\eps\in(0, 1/10]$ and $\delta\in(0,1/2)$.
  Let $s = \lfloor \delta \ln(m)/\,250\eps^2 p \rfloor$.
  Assume that $s \le m^{1/2-\delta}$,
  that $n \le m^{1/\delta}$,
  and that $m$ is sufficiently large (exceeding some constant that depends only on $\delta$).
  Then

 \medskip\noindent{\emph{~(i)}}
 {\(\displaystyle\Pr_B[\Vmax B \ge (1-3\eps)p] ~\le~ 1/n^{2s}\),}

 \medskip\noindent{\emph{(ii)}}
 {\(\displaystyle\Pr_B[\Vmin B\, \le (1+3\eps)p] ~\le~ 1/n^{2s}\).}
\end{lemma}
\begin{proof}
  We check the technical assumptions necessary to apply Lemma~\ref{lemma:skewed},
  and check that the upper bound from that lemma
  implies the upper bound claimed in this lemma.
  By inspection of $s$, the condition $\eps^2ps \ge 1$ of Lemma~\ref{lemma:skewed}
  is satisfied for $m = \exp(\Omega(1/\delta))$.  
  To finish, we show, for this $s$ and $t=m\exp({-250\eps^2ps})$ from Lemma~\ref{lemma:skewed},
  that the upper bound $2s\exp({-\eps^2t p/15})$ from that lemma is at most $1/n^{2s}$ (for large enough $m$).

  \smallskip
  \noindent
  If $s=0$, the corollary is trivial, so assume without loss of generality that $s\ge 1$.
  Then\noindent
\begin{align}
  \LINE{L1}
  {\delta\ln(m)/\,500\eps^2 p}{\,\le\, s\,\le\, \delta \ln(m)/\,250 \eps^2 p}
  {By the choice of $s$ and $s\ge 1$.}
  \\
  \LINE{L2}
  {t}{\,\ge\, m^{1-\delta}}
  {By substituting the right-hand side (RHS) of~\eqref{L1}
  for $s$ in the definition of $t$ and simplifying.}
\\
\LINE{L3}
{m^{1-2\delta}}{\,\ge\, s^2}
{Squaring both sides of assumption $s\le m^{1/2-\delta}$.}
\\
\LINE{L4}
{m^{1-\delta}}{\,\ge\, 50\times 500\, m^{1 - 2\delta}/\delta^{2} }
{For sufficiently large $m$ (depending only on $\delta$).} 
 \\
 \LINE{L5}
 {m^{1-\delta}}{\,\ge\, 50\times 500\, s^2/\delta^{2} }
 {Substituting RHS of~\eqref{L3}
   for $m^{1-2\delta}$ in~\eqref{L4}.}
\\
\LINE{L6}
{t}{\,\ge\, 50\times 500\, s^2/\delta^{2}}
{By transitivity on~\eqref{L2} and~\eqref{L5}.}
\\
\LINE{L7}
{t}{\,\ge\, 50 s \ln(m)/ \delta \eps^2p}
{By substituting the left-hand side (LHS) of~\eqref{L1}
  for one $s$ in~\eqref{L6} and simplifying.}
\\
\LINEL{L8}
{t}{\,\ge\, 15[2 s \ln(m)/\delta+\ln 2s]/\eps^2p}
{By~\eqref{L7} and calculation, for large enough $m$.}
\\
\LINE{L9}
{t}{\begin{array}[t]{ll}
       \,\ge\, 15[2 s\ln n+\ln 2s]/\eps^2p
       \\[2pt]\,=\, 15 \ln(2s n^{2s}) / \eps^2 p
    \end{array}}
{By~\eqref{L8} and assumption $n\le m^{1/\delta}$,
   that is, $\ln n \le \ln(m)/\delta$.}
\\
\LINE{L10}
{\eps^2 t p/15}{\,\ge\, \ln(2 s n^{2s})}
{Rearranging~\eqref{L9}.}
\\
\LINE{L11}
{2s\exp({-\eps^2 t p/15})}{\,\le\, 1/n^{2s} \notag}
{By~\eqref{L10}, taking exponentials and rearranging.}
\end{align}
\end{proof}

This concludes the utility lemmas.   Next we state and prove Theorem~\ref{thm:game}.

\begin{figure}\centering\small
  \begin{tabular}{@{}l@{}r@{}}
    \parbox[t]{0.44\textwidth}{
      \centerline{\emph{\underline{given parameters}}}\smallskip
      \begin{tabular}[t]{@{}r@{ = }l@{}} 
        $m$ & number of constraints (rows)\\
        $n$ & number of variables (columns) \\
        $\delta$ & constant in (0,1/2)\\
        $\rho$ & approximate width, $\rho \ge 2$\\
        $p$ & $1/\rho = \Pr[A_{ij}=1]$, $p\in(0,1/2)$
      \end{tabular}
    }
    &
    \parbox[t]{0.56\textwidth}{
      \centerline{\emph{\underline{determined from given}}}\smallskip
      \begin{tabular}[t]{@{}r@{ = }l@{}} 
        $A$~~& $m\times n$, 0/1 matrix w/ i.i.d.~entries\\
        {\sf packing}$(A)$& \emph{``find $x\in \Delta^n$ minimizing $\max_i Ax$'' } \\
        {\sf covering}$(A)$& \emph{``find $x\in \Delta^n$ maximizing $\min_i Ax$''} \\
        $\Vmin{A}$& value of game $A$ if {\sc min} plays cols\\
        $\Vmax{A}$& value of game $A$ if {\sc max} plays cols
      \end{tabular}}
  \end{tabular}
  \caption{notation for Theorem~\ref{thm:game} and Theorem~\ref{thm:main}.}\label{fig:key}
\end{figure}

\newsavebox\thmGame
\savebox{\thmGame}{\parbox{\textwidth}{\setlength{\parindent}{1.5em}
    \begin{theorem}[support-size bound]\label{thm:game}
      For every constant $\delta\in(0,1/2)$, 
      there exists constant $k_\delta>0$ such that the following holds.
      Fix arbitrary integers $m,n > k_\delta$ and arbitrary $p\in(0,1/2)$.
      Let $\eps_0$ be such that  $p^{-1}\,\eps^{-2}_0 \ln(m) = \min(m^{1/2-\delta}, n/9)$.
      Assume $n \le m^{1/\delta}$ and $\eps_0 \le 1/10$.
      Let $A$ be a random $m\times n$ 0/1 matrix with i.i.d.~entries, 
      where each entry $A_{ij}$ is 1 with probability $p$.
      Then, with probability $1-O(1/m^2)$,
      \begin{enumerate}
      \item both $\Vmax A$ and $\Vmin A$ lie in the interval $[(1-\eps_0)p,\, (1+\eps_0)p]$, and
      \item for all $\eps\in[\eps_0, 1/10]$,
        every $(1+\eps)$-approximate mixed strategy for the column player (as {\sc Min} or as {\sc Max})
        has support of size at least
        $\delta\, p^{-1}\,\eps^{-2}\ln(m)\,/\,1000$.
      \end{enumerate}
    \end{theorem}
}}
\noindent\usebox{\thmGame}

\begin{proof}
All probabilities in the proof
are with respect to the random choice of $A$.

\smallskip

\noindent
\emph{Part 1, bounds on $\Vmin A$ and $\Vmax A$:}  
By the naive bound (Lemma~\ref{lemma:naive}), the probability that
either $\Vmin A$ or $\Vmax A$ fall either before or after the interval is at most
\[ 2\, m\,\exp({-n\,\eps_0^2\,p /3}) + 2\, n\,\exp({-m\,\eps_0^2\,p /3}).\]
The first of the two terms is at most $2/m^2$ 
because the definition of $\eps_0$ implies $\eps_0^2\, p\, \ge \, 9\ln(m)/n$.

Likewise, the second of the first two terms is at most $2/m^2$,
because, using the definition of $\eps_0$ again, $\eps_0^2\, p\,\ge m^{1/2-\delta}$, so
\[
m\, \eps_0^2\, p \,\ge\, m^{1/2+ \delta} \,\ge\, m^{1/2}\,\ge\, 3\,\ln n~+~ 6\ln m
\]
(using that $m$ is large enough so that
$0.9 \sqrt m \ge 3\delta^{-1} \ln m \ge 3 \ln n$ and $0.1 \sqrt m \ge 6 \ln m$).

\smallskip 
\noindent
\emph{Part 2.}
Define r.v.~$\Smin{\eps}$ to be the minimum support size of any
mixed strategy that achieves value $(1+\eps)p$ or less for {\sc min}
as the column player of $A$.
Analogously let $\Smax \eps$ be the minimum support size
that achieves value at least $(1-\eps)p$ for {\sc max} as the column player.

Next we use the skewed-game lemma (Lemma~\ref{lemma:skewed})
and the union bound to bound
the probability that either player (playing the columns of $A$) has a good strategy with small support.
Let $\ell(\eps) = \delta\, p^{-1} \eps^{-2} \ln(m)/1000$ denote the desired lower bound on the support size
for a given $\eps$.

\begin{thmObservation}\label{obs:skewed2}
  Let $\eps\in [\eps_0,1/10]$.
  Let $s = \lfloor 4\,\ell(\eps) \rfloor = \lfloor \delta\ln(m)/\,250\eps^2 p \rfloor$.  
  If $s\le m^{1/2-\delta}$ and $n \le m^{1/\delta}$, then

  \medskip\noindent{\emph{~(i)}}
 $\Pr_A[S^*_{\max}(3\eps) \le s] ~\le~ 1/n^{s}$, and

 \medskip\noindent{\emph{(ii)}}
 $\Pr_A[S^*_{\min}(3\eps) \le s] ~\le~ 1/n^{s}$.
\end{thmObservation}

\begin{figure}
  \centering
  \includegraphics[width=1.5in]{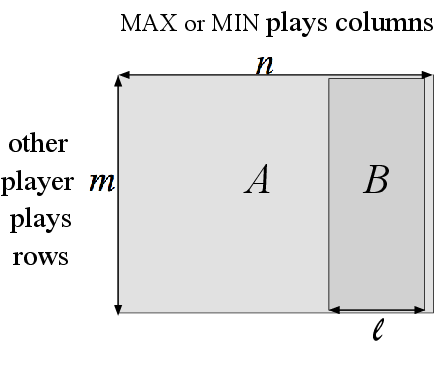}
  \caption{The matrix $A$, and one submatrix $B$ consisting of $\ell(\eps)$ columns.
  }\label{fig:matrices}
\end{figure}

(Note that $\Smax{3\eps} \le s$ iff {\sc max} can get value $(1-3\eps)p$ or more using at most $s$ columns.)
\smallskip

\begin{proof}
  \emph{(i)}
  If {\sc max} has a mixed strategy with support of size $s$
  that has value at least $(1-3\eps)p$,
  then $A$ has an $m\times s$ submatrix $B$ with $\Vmax B \ge (1-3\eps)p$.

  Consider all $n \choose s$ possible such submatrices $B$.
  By Lemma~\ref{lemma:skewed2}, given any one of these submatrices $B$,
  the probability of $\Vmax B\ge (1-3\eps)p$ is at most $1/n^{2s}$.
  Thus, by the union bound, the probability that \emph{any} such submatrix $B$ of $A$
  has $\Vmax B\ge(1-3\eps)p$ is at most ${n\choose s}/n^{2s} \le 1/n^{s}$.

  \smallskip
  The proof for \emph{(ii)} is essentially the same.
\end{proof}


\medskip
Observation~\ref{obs:skewed2} bounds the probability of failure for a \emph{single} given $\eps$.
We want to show that \whp the bound holds for all $\eps\in(0, 1/10]$ simultaneously.
We start by considering a sequence $Q$ of geometrically increasing $\eps$ values:
$Q=\{ {2^i}\eps_0 : i=0,1,2,\ldots,\lfloor\log_{2} (0.1/\eps_0)\rfloor\}$.

The maximum $\eps$ in $Q$ is just less than $0.1$.

\begin{thmObservation}\label{obs:Q}
With probability $1-O(1/m^2)$, for all $\eps\in Q$,
support of size $4\ell(\eps)$ is necessary
for {\sc max} to achieve value $(1-3\eps)p$
or
for {\sc min} to achieve value $(1+3\eps)p$.
Specifically, for $n$ and $m$ large enough (as a function of $\delta$),
with probability $1-O(1/m^2)$,
for all $\eps\in Q$,~\,$\Smax{3\eps}> 4\,\ell(\eps)$ and $\Smin{3\eps}> 4\,\ell(\eps)$.
\end{thmObservation}
\begin{proof}
By Observation~\ref{obs:skewed2},
for every $\eps$ in the set $Q$,
the probability of the event $\Smax{3\eps}\le 4\,\ell(\eps)$ is at most $1/n^{\lfloor 4 \ell(\eps)\rfloor}$.
By the union bound,
the probability that
there exists an $\eps\in Q$ with $\Smax{3\eps}\le 4\,\ell(\eps)$ is at most
\(
\sum_{\eps \in Q} 1/n^{\lfloor 4 \ell(\eps)\rfloor}.
\)
Using that $\ell(2^i \eps_0) = \ell(\eps_0)/4^i$ for $i \ge 0$,
and the definition of $Q$,
this sum is at most
\(\sum_{i=0}^\infty 1/n^{\lfloor 4^i \ell(0.1)\rfloor}\).  
The terms in this sum decrease super-geometrically,
so the sum is proportional to its first term,
which is at most $1/n^{\delta \ln(m)/ 10 p\, - 1}$,
which is $O(1/m^2)$ as long as $n$ and $m$ are large enough (as a function of $\delta$).

The proof for {\sc min} is similar.
\end{proof}

\medskip
To complete the proof of Theorem~\ref{thm:game},
we extend the previous observation to all $\eps\in[\eps_0, 1/10]$:
\begin{thmObservation}\label{obs:complete}
With probability $1-O(1/m^2)$, 
for all $\eps\in [\eps_0,1/10]$,
support of size $\ell(\eps)$ is necessary
for {\sc max} to achieve value $(1-3\eps)p$
and
for {\sc min} to achieve value $(1+3\eps)p$.
Specifically, for $n$ and $m$ large enough (as a function of $\delta$),
with probability $1-O(1/m^2)$, 
for all $\eps\in [\eps_0,1/10]$,~\,$\Smax{3\eps}> \,\ell(\eps)$ and $\Smin{3\eps}> \,\ell(\eps)$.
\end{thmObservation}
\begin{proof}
We show that if the event in Observation~\ref{obs:Q} happens,
then the event desired above happens.
Assume the former event happens,
i.e., $\forall \eps'\in Q,~ S^*_{\max}(3\eps') > 4\,\ell(\eps')$.

Now consider any $\eps\in [\eps_0, 1/10]$.
By the choice of $Q$,
there is some $\eps'\in Q$ such that $\eps \in (\eps'/2, \eps']$.
Then we have
\begin{romannum}
\item $\Smax{3\eps} \ge \Smax{3\eps'}$ (since $\Smax{\cdot}$ is monotone decreasing and $\eps\le \eps'$),
\item $\Smax{3\eps'} \ge 4\,\ell(\eps')$ (since $\eps'\in Q$), and
\item $\ell(\eps') > \ell(\eps)/4$ (by the definition of $\ell(\cdot)$ and $\eps > \eps'/2$).
\end{romannum}
By transitivity, we conclude that $\Smax{3\eps} > \ell(\eps)$
for all $\eps\in[\eps_0,1/10]$.

The proof for $\Smin{\eps}$ is similar.
\end{proof}

We now finish the proof of Theorem~\ref{thm:game}, Part (ii).
From Part 1 of the theorem, with probability $1-O(1/m^2)$, for all $\eps\in[\eps_0,1/10]$,
to achieve $(1+\eps)$-approximation,
{\sc max} must achieve absolute value at least $(1-\eps_0) p / (1+\eps) \ge (1-3\eps)p$.
By Observation~\ref{obs:complete}, with probability $1-O(1/m^2)$, for all $\eps\in[\eps_0,1/10]$,
support size at least $\ell(\eps)$ is needed for {\sc max} to achieve this absolute value.
By the union bound,
with probability $1-O(2/m^2)$, every $(1+\eps)$-approximate strategy 
for {\sc max} has support size at least $\ell(\eps)$.
By a similar argument
(using $(1+\eps_0)(1+\eps)p \le (1+3\eps)p$),
with probability $1-O(1/m^2)$, every $(1+\eps)$-approximate strategy
for {\sc min} also has support size at least $\ell(\eps)$.
This completes the proof of Theorem~\ref{thm:game}.
\end{proof}

Before we prove Theorem~\ref{thm:main}, 
we observe that Corollary~\ref{cor:game}
is indeed just a simplified (and somewhat weaker) statement of Theorem~\ref{thm:game}:

\begin{proof}\emph{(Corollary~\ref{cor:game})}
Fix any $\delta$, $m$, $n$ and $p$ as in the corollary. 
(Take $k_\delta$ in the corollary to be the same as in the theorem.)
Assume without loss of generality that $n\le m^{1/\delta}$.
(Otherwise, decrease $n$ to $n' = \lfloor m^{1/\delta} \rfloor \ge m^2$,
apply the corollary to get a game with $m\times n'$ payoff matrix $A'$,
then duplicate any of the columns $n-n'$ times to get an equivalent $m\times n$ 
game with the desired properties.)

Redefine $\ell(\eps) = p^{-1}\eps^{-2} \log m$.
Fix $\eps_0$ as in the theorem.
The choice of $\eps_0$ implies $\ell(\eps_0) = \Theta(\min(m^{1/2-\delta}, n))$,
so the support bound desired for the corollary is equivalent to

\smallskip

\noindent
\emph{\small $\forall\,\eps\in (0, 1/10]$,
    any $(1+\eps)$-approximate mixed strategy has support size
    $\Omega_\delta\big(\ell\big(\max(\eps_0, \eps)\big)\big)$.}

\smallskip

Assume without loss of generality that $\eps_0 \le 1/10$.
(If $\eps_0 > 1/10$, raise $p$ until the corresponding $\eps_0$ decreases to $1/10$;
the corollary for the smaller $p$ follows from the corollary for the larger $p$,
because in both cases the lower bound in question is the same:
$\Omega_\delta(\min(m^{1/2-\delta}, n))$, $\forall \eps\in (0,1/10)$.)

Now we have $\eps_0 \le 1/10$ and $n \le m^{1/\delta}$.
Applying Theorem~\ref{thm:game}, there are (many) $m\times n$ zero-sum matrix games with value $\Omega(p)$
such that, for all $\eps\in [\eps_0, 1/10]$, any $(1+\eps)$-approximate strategy for the column
player requires support of size at least $\Omega_\delta(\ell(\eps))$.
To finish, note that, for the remaining $\eps \in (0, \eps_0]$, any $(1+\eps)$-approximate strategy 
is also a $(1+\eps_0)$-approximate strategy,
so must have support size at least $\Omega_\delta(\ell(\eps_0))=\Omega_\delta(\min(m^{1/2-\delta}, n))$,
proving the corollary.
\end{proof}

\section{Theorem~\ref{thm:main} (iteration bound)}
\label{sec:main}

Before we state and prove Theorem~\ref{thm:main}, we prove two utility lemmas.
The first says that the output $\hat x$ of any Dantzig-Wolfe-type algorithm has to be a convex combination
of the vectors output by the oracle.

\begin{lemma}\label{lemma:inP}
  Suppose that a deterministic Dantzig-Wolfe-type algorithm, 
  given some input $(A,b,X_P)$, returns a solution $\hat x\in P$.
  Then $\hat x$ must be a convex combination of the outputs
  returned by the oracle $X_P$ during the computation.
  The same holds if the algorithm is randomized (and has zero probability of error).
\end{lemma}
\begin{proof}
  First we consider the deterministic case.  
  Let $Q$ denote the set of oracle inputs generated by the algorithm on input $(A,b,X_P)$.
  Define polyhedron $P'\subseteq P$ 
  to be the convex hull of the vectors output by $X_P$ during the algorithm.
  That is $P'$ is the polyhedron whose vertices are $\{X_{P}(q) : q\in Q\}$.
  Suppose for contradiction that $\hat x\not\in P'$
  and consider the modified input $(A,b,P')$, with polyhedron $P'$ instead of $P$.
  Define the oracle $X'_{P'}$ for the polyhedron $P'$ 
  such that $X'_{P'}(q)$ outputs a minimizer of $q\tran x$ among $x\in P'$.
  For $q\in Q$, break any ties among the minimizers for $q$ by choosing $X'_{P'}(q) = X_P(q)$.
  This $X'_{P'}$ optimizes correctly over $P'$.
  Observe that it also has the following key property:
  \emph{Let $q\in Q$ be any input that the algorithm gave to oracle $X_P$ on input $(A,b,X_P)$.
    Then, on input $q$, oracle $X'_{P'}$ gives the same output, $X_P(q)$, that $X_P$ did.}

  Consider rerunning the Dantzig-Wolfe-type algorithm, 
  this time on the input $(A,b,X'_{P'})$.
  The Dantzig-Wolfe-type algorithm is deterministic,
  and, as observed above, $X_P(q) = X'_{P'}(q)$ 
  for all inputs $q\in Q$ that the algorithm gave to the oracle
  when the algorithm ran on input $(A,b,X_P)$.
  Recall that the algorithm interacts with the polyhedron only via the oracle ($X_P$ or $X'_{P'}$).
  By induction on the number of queries, when run on $(A,b,X'_{P'})$,
  the algorithm behaves the same
  --- that is, it makes the same sequence of queries and computes the same final answer $\hat x$ ---
  as it did when run on $(A,b,X'_{P'})$.
  But this is an incorrect output, as $\hat x$ is not in the polyhedron $P'$ for the second input.
  This proves the lemma for the deterministic case.

  \smallskip
  Now consider running any (error-free) randomized Dantzig-Wolfe-type algorithm on $(A,b,X_P)$.
  Suppose for contradiction that the algorithm has non-zero probability
  of producing an output $\hat x$ that is not a convex combination 
  of the oracle outputs made during the run.
  Fix any such outcome that has positive probability, say $p'>0$.
  Let $Q$, $P'$ and $X'_{P'}$ be as in the proof above,
  and consider running the algorithm on input $(A,b,X'_{P'})$.
  With probability at least $p'$, 
  the algorithm will make the same random choices that it made in the first run.
  When this happens, then (as in the proof for the deterministic case)
  it returns the same vector $\hat x$, which is (just as for the deterministic case) an error,
  because $\hat x \not\in P'$.
  Hence, the algorithm has positive probability of error on input $(A,b,X'_{P'})$.
\end{proof}

The next lemma is a convenient restatement of Theorem~\ref{thm:game} 
in terms of {\sf packing}$(A)$ and {\sf covering}$(A)$.

\begin{lemma}\label{lemma:middle}
  For every constant $\delta \in (0,1/2)$, 
  there exists constant $k_\delta>0$ such that the following holds.
  Fix any integers $m, n > k_\delta$ and any desired width $\rho\ge 2$.  
  Let $\eps_0$ be such that $\rho\,\eps^{-2}_0 \ln(m) = \min(m^{1/2-\delta}, n/9)$.
  Assume $n\le m^{1/\delta}$ and $\eps_0 \le 1/10$.
  Let $A$ be a random $m\times n$ 0/1 matrix with i.i.d.~entries, 
  where each entry $A_{ij}$ is 1 with probability $p=1/\rho$.
  Then, with probability $1-O(1/m^2)$, for both {\sf packing}$(A)$ and {\sf covering}$(A)$,
  \begin{remunerate}
  \item the instance has width at most $2\rho$,
  \item for all $\eps \in [\eps_0, 1/10]$,
    every $(1+\eps)$-approximate solution
    has support size at least
    \[\delta\, \rho\,\eps^{-2}\ln(m)\,/\,1000.\]
  \end{remunerate}
\end{lemma}
\begin{proof}
  Note that we take $p=1/\rho$.

  The $(1+\eps)$-approximate solutions to {\sf packing}$(A)$ and {\sf covering}$(A)$ are,
  respectively, the $(1+\eps)$-approximate mixed strategies for {\sc min} and {\sc max} 
  (as the column player of the game with payoff matrix $A$).
  Thus, Part 2 of Theorem~\ref{thm:game} implies Part 2 of the lemma.

  Regarding Part 1 of the lemma, 
  suppose that Part 1 of the theorem holds,
  so $\Vmin{A}$ and $\Vmax{A}$ are both at least $(1-\eps_0)p$.
  By definition of {\sf packing}$(A)$, each $b_i$ is $\Vmin{A}$,
  so the width is $\max_{x,i} A_i x / \Vmin{A}$,
  where $x$ ranges over the simplex.
  Since $A$ is a 0/1 matrix and $\sum_j x_j = 1$
  (and $A$ is not all zeros, as $\Vmin{A} \ge (1-\eps_0)p > 0$)
  we have $\max_{x,i} A_i x = \max_{ij} A_{ij} = 1$,
  so the width is $1 / \Vmin{A} \le 2/p = 2 \rho$.
  The same argument shows that {\sf covering}$(A)$ 
  has width $1/\Vmax{A} \le 2\rho$.
\end{proof}

Next we state and prove Theorem~\ref{thm:main}.

\newsavebox\thmMain
\savebox{\thmMain}{\parbox{\textwidth}{\setlength{\parindent}{1.5em}
    \begin{theorem}[iteration bound]\label{thm:main}
      For every constant $\delta \in (0,1/2)$, 
      there exists constant $k_\delta>0$ such that the following holds.
      Fix arbitrary integers $m, n > k_\delta$ and an arbitrary desired width $\rho\ge 2$.  
      Let $\eps_0$ be such that $\rho\,\eps^{-2}_0 \ln(m) = \min(m^{1/2-\delta}, n/9)$.
      Assume $n\le m^{1/\delta}$ and $\eps_0 \le 1/10$.
      Let $A$ be a random $m\times n$ 0/1 matrix with i.i.d.~entries, 
      where each entry $A_{ij}$ is 1 with probability $p=1/\rho$.
      Then, with probability $1-O(1/m^2)$, for both {\sf packing}$(A)$ and {\sf covering}$(A)$,
      \begin{remunerate}
      \item the instance has width at most $2\rho$,
      \item for all $\eps \in [\eps_0, 1/10]$,
        all deterministic Dantzig-Wolfe-type algorithms,
        and all Las-Vegas-style randomized Dantzig-Wolfe-type algorithms, must make at 
least
        \(\delta\, \rho\,\eps^{-2}\ln(m)\,/\,1000\) 
        iterations to find a $(1+\eps)$-approximate solution.
      \end{remunerate}
    \end{theorem}
}}\noindent\usebox{\thmMain}

\begin{proof}
  Fix any values of the parameters $\delta,m,n,\rho,p=1/\rho$.
  Let $A$ be as described.
  Assume the events 1 and 2 in Lemma~\ref{lemma:middle} happen for $A$
  (as they do with probability $1-O(1/m^2)$).

  \noindent
  \emph{ Part 1.} Part 1 of the theorem is immediate from event 1 of Lemma~\ref{lemma:middle}.

  \noindent
  \emph{ Part 2.}
  Let $e_j$ denote the $j$th standard basis vector for $\R^n$,
  that is, the vector that is $1$ in the $j$th coordinate and zero elsewhere,
  so that the set of vertices of $\Delta^n$ is $\{e_j : j\in[n]\}$.

  Fix any oracle $X_n$ whose output $X_n(q)$ for each input $q$ 
  is some vertex $e_j$ of $\Delta^n$ (one minimizing $q\tran e_j$; breaking ties consistently).
  For any $\eps\in[\eps_0,1/10]$,
  run the (deterministic or randomized) Dantzig-Wolfe-type algorithm on input $(A,b,X_n)$.
  Let $\hat x$ be the $(1+\eps)$-approximate solution it returns.

  By Lemma~\ref{lemma:inP}, $\hat x$ is a convex combination of the vectors returned by the oracle.
  Each such vector has just one non-zero coordinate.
  Thus, the number of non-zero coordinates in $\hat x$ 
  is at most the number of iterations made by the algorithm.
  On the other hand, $\hat x$ is a $(1+\eps)$-approximate solution,
  so  by event (2) of Lemma~\ref{lemma:middle},
  the number of non-zero coordinates is at least the desired lower bound
  \(\delta\, \rho\,\eps^{-2}\ln(m)\,/\,1000\).
\end{proof}

Finally, we observe that Corollary~\ref{cor:main}
is indeed just a simplified and somewhat weaker statement of Theorem~\ref{thm:main}.
The proof is identical to 
the proof that Corollary~\ref{cor:game} 
follows from Theorem~\ref{thm:game}.

\begin{proof}\emph{(Corollary~\ref{cor:main})}
Fix any $\delta$, $m$, $n$ and $p$ as in the corollary. 
(Take $k_\delta$ in the corollary to be the same as in the theorem.)
Assume without loss of generality that $n\le m^{1/\delta}$.
(Otherwise, decrease $n$ to $n' = \lfloor m^{1/\delta} \rfloor \ge m^2$,
apply the corollary to get $m\times n'$ packing or covering instances,
then duplicate any of the columns $n-n'$ times to get equivalent $m\times n$ 
instances with the desired properties.)

Let $\ell(\eps) = p^{-1}\eps^{-2} \log m$.
Fix $\eps_0$ as in the theorem.
As $\ell(\eps_0) = \Theta(\min(m^{1/2-\delta}, n))$,
the support bound desired for the corollary is equivalent to

\smallskip

\noindent
  \emph{$\forall \eps\in (0, 1/10]$,
    any $(1+\eps)$-approximate solution has support size
    $\Omega_\delta\big(\ell\big(\max(\eps_0, \eps)\big)\big)$.}

\smallskip

Assume without loss of generality that $\eps_0 \le 1/10$.
(If $\eps_0 > 1/10$, lower $\rho$ until the corresponding $\eps_0$ decreases to $1/10$;
the corollary for the larger $\rho$ follows from the corollary for the smaller $\rho$,
because in both cases the lower bound in question is the same:
$\Omega_\delta(\min(m^{1/2-\delta}, n))$, $\forall \eps\in (0,1/10)$.)

Now we have $\eps_0 \le 1/10$ and $n \le m^{1/\delta}$.
Applying Theorem~\ref{thm:main}, there are (many) $m\times n$ packing/covering instances 
with width $O(\rho)$
such that, for all $\eps\in [\eps_0, 1/10]$, every $(1+\eps)$-approximate solution
has support of size at least $\Omega_\delta(\ell(\eps))$.
To finish, note that, for the remaining $\eps \in (0, \eps_0]$, any $(1+\eps)$-approximate solution
is also a $(1+\eps_0)$-approximate solution,
so must have support size at least $\Omega_\delta(\ell(\eps_0))=\Omega_\delta(\min(m^{1/2-\delta}, n))$,
proving the corollary.
\end{proof}

\section*{Acknowledgements}

The authors thank editor Robert Krauthgamer
and several anonymous referees for feedback that
helped improve the presentation substantially,
and Xinhua Zhang
(Machine Learning Research Group, National ICT Australia)
for pointing out that the argument in the conference version of this paper
does not require $n \approx \Theta(m^{1/2})$~\cite{KleinY99}.

{
\bibliographystyle{siam}
\bibliography{all}
}



\section*{Appendix --- Lemma~\ref{lemma:tightness} (tightness of Chernoff bound)}


Here is the proof of Lemma~\ref{lemma:tightness} ---
that a standard Chernoff bound 
is tight up to constant factors in the exponent for a particular range of the parameters.
(In particular, whenever the variables are 0 or 1, and 1 with probability 1/2 or less, 
and $\epsilon\in(0,1/2)$, and the Chernoff upper bound is less than a particular constant.)
First we prove the following useful inequality:
\begin{lemma}\label{lemma:stirling}
 If $1\le \ell\le k-1$, then
  \(\displaystyle {k \choose \ell} ~\ge~ \frac{1}{e\sqrt{2\pi\ell}} {\Big(\frac{k}{\ell}\Big)}^{\ell} {\Big(\frac{k}{k-\ell}\Big)}^{k-\ell}\)
\end{lemma}
\begin{proof}
  By Stirling's approximation, 
  \(i!=\sqrt{2\pi i}{(i/e)}^i e^\lambda\) where \(\lambda\in[\frac{1}{12i+1},\frac{1}{12i}]\).
  Thus, $k\choose \ell$ is
  \begin{eqnarray*}
    \frac{k!}{\ell! (k-\ell)!}
    &\ge&
    \frac{\sqrt{2\pi k}\,{(\frac{k}{e})}^k}
    {  \sqrt{2\pi \ell}\,{(\frac{\ell}{e})}^\ell
      ~~\sqrt{2\pi (k-\ell)}\,{(\frac{k-\ell}{e})}^{k-\ell} }
      \exp\Big(\frac{1}{12k+1} - \frac{1}{12\ell} - \frac{1}{12(k-\ell)}\Big)
    \\
   & \ge  &
    \frac{1}{\sqrt{2\pi\ell}} {\Big(\frac{k}{\ell}\Big)}^{\ell} {\Big(\frac{k}{k-\ell}\Big)}^{k-\ell}e^{-1}.
 \end{eqnarray*}
\vspace*{-2.5em}

\end{proof}

\newcounter{saveTheorem}
\setcounter{saveTheorem}{\value{theorem}}
\newcounter{saveSection}
\setcounter{saveSection}{\value{section}}
\setcounter{section}{\value{tightnesssection}}
\setcounter{theorem}{\value{tightnesscounter}}

\begin{lemma}[tightness of Chernoff bound]
  \chernoffTight
\end{lemma}
\begin{proof}
  \emph{Part (i).}
Without loss of generality assume each 0/1 random variable in the sum $X$
is 1 with probability \emph{exactly} $p$.
Note $\Pr[X\le (1-\eps)p]$ equals the sum $\sum_{i = 0}^{\lfloor(1-\eps)pk\rfloor} \Pr[X=i/k]$,
and $\Pr[X=i/k] = {k \choose i} p^i {(1-p)}^{k-i}$.
 
  \medskip

  Fix $\ell = \lfloor(1-2\eps)pk\rfloor+1$.
  The terms in the sum are increasing,
  so the terms with index $i\ge\ell$
  each have value at least $\Pr[X=\ell/k]$,
  so their sum has total value at least
  $(\eps pk - 2) \Pr[X=\ell/k]$.
 To complete the proof, we show that 
 \[(\eps pk - 2) \Pr[X=\ell/k] ~\ge~ \exp({-9\eps^2 pk}).\]

 The assumptions $\eps^2pk\ge 3$ and $\eps\le 1/2$
  give $\eps pk \ge 6$,
  so the left-hand side above is at least $\frac{2}{3}\eps pk\, {k \choose \ell} p^\ell {(1-p)}^{k-\ell}$.
  Using Lemma~\ref{lemma:stirling} 
  to bound $k\choose \ell$,
  this is in turn at least $A\, B$
  where
 $A =  \frac{2}{3e}\eps p k/ \sqrt{2\pi \ell}$
 and
  \(
  B=
  {\big(\frac{k}{\ell}\big)}^\ell 
  {\big(\frac{k}{k-\ell}\big)}^{k-\ell}
  p^\ell {(1-p)}^{k-\ell}.
  \)
  To finish we show $A\ge \exp(-\eps^2pk)$ and $B \ge \exp(-8\eps^2 pk)$.
 
  \begin{lemmaObservation}\label{observation:two}
    $A \ge \exp({-\eps^2 pk})$
  \end{lemmaObservation}
  \begin{proof}
    The assumption $\eps^2pk \ge 3$
    implies ${\exp(-\eps^2pk)}{\,\le\, \exp(-3) \le 0.04}$.
    To finish we show $A\ge 0.1$:
    \begin{align}
      \LINE{M1}
      {12}{\,\le\, pk}
      {By assumptions $\eps^2 pk \ge 3$ and $\eps\le 1/2$.}
      \\
      \LINE{M2}
      {\ell}{\,\le\, 1.1 pk}
      {From~\eqref{M1} and $\ell \le pk + 1$ (from $\ell$'s definition).}
      \\
      \LINE{M3}
      {A}{\,\ge\,\frac{2}{3e} \eps \sqrt{p k / 2.2\pi}}
      {Using~\eqref{M2} to substitute for $\ell$ in definition of $A$.}
      \\
      \LINE{M4}
      {A}{\,\ge\,\frac{2}{3e}\sqrt{3/2.2\pi} \,\ge\, 0.1}
      {From~\eqref{M3} and $\eps\sqrt{ pk} \ge \sqrt 3$
        (from $\eps^2 pk \ge 3$).}\notag
    \end{align}
  \end{proof}
  
  \begin{lemmaObservation}\label{observation:three}
    $B\ge \exp({-8\eps^2 pk})$.
  \end{lemmaObservation}
  \begin{proof}
    Fix $\delta$ such that $\ell=(1-\delta)pk$.  
    The choice of $\ell$ implies $\delta\le 2\eps$,
    so
    the observation will hold as long as $B \ge \exp(-2\delta^2pk)$.
    Taking each side of this latter inequality to the power $-1/\ell$ and simplifying,
    it is equivalent to
    \[
    \frac{\ell}{p k}  {\Big(\frac{k-\ell}{(1-p) k}\Big)}^{k/\ell-1}
    ~\le~
    \exp\Big(\frac{2\delta^2 pk}{\ell}\Big).
    \]
    Substituting $\ell= (1-\delta)pk$ and simplifying, it is equivalent to
    \[
    (1-\delta)  {\Big(1+\frac{\delta p}{1-p}\Big)}^{\displaystyle \frac{1}{(1-\delta)p}-1}
    ~\le~
    \exp\Big(\frac{2\delta^2}{1-\delta}\Big).
    \]
    Taking the logarithm of both sides and using $\ln(1+z)\le z$ twice,
    it will hold as long as
    \[
    -\delta\, +\,\frac{\delta p}{1-p}\Big(\frac{1}{(1-\delta)p}-1\Big)
    ~\le~
    \frac{2\delta^2}{1-\delta}.
    \]
    The left-hand side above simplifies to $\delta^2/\,(1-p)(1-\delta)$,
    which is less than $2\delta^2/(1-\delta)$ because $p\le 1/2$.
  \end{proof}
  
\medskip

Observations~\ref{observation:two} and~\ref{observation:three}
imply $A B \ge \exp({-\eps^2pk})\exp({- 8\eps^2pk})$.
This implies Part (i) of Lemma~\ref{lemma:tightness}.

  \medskip

  \noindent\emph{Part (ii).}
  Without loss of generality assume each random variable is $1$ with probability exactly $p$.

  Note $\Pr[X\ge (1+\eps)p] = \sum_{i = \lceil(1-\eps)pk\rceil}^n \Pr[X=i/k]$.
  Fix $\hat\ell = \lceil (1+2\eps)pk \rceil - 1$.

  \smallskip
  The last $\eps pk$ terms in the sum
  total at least  $(\eps pk-2)\Pr[X=\hat\ell/k]$,
  which is at least $\exp({-9\eps^2 pk})$.
  (The proof of that is the same as for (i), except with $\ell$ replaced by $\hat\ell$
  and $\delta$ replaced by $-\hat\delta$ such that $\hat\ell = (1+\hat\delta)pk$.)
\end{proof}
\setcounter{theorem}{\value{saveTheorem}}
\setcounter{section}{\value{saveSection}}

\end{document}